\documentclass[conference]{IEEEtran}
\IEEEoverridecommandlockouts
\usepackage{cite}
\usepackage{amsmath,amssymb,amsfonts}
\usepackage{algorithmic}
\usepackage{graphicx}
\usepackage{textcomp}
\usepackage{xcolor}
\usepackage{cite}
\usepackage{balance}
\usepackage{bm}
\usepackage{algorithm}
\usepackage{algorithmic}
\usepackage{caption}
\usepackage{subcaption}

\usepackage[letterpaper, left=0.68in, right=0.68in, bottom=1.05in, top=0.7in]{geometry}

\begin{document}
	
\bstctlcite{IEEEexample:BSTcontrol}

\title{
	{
		Energy-Efficient Throughput Maximization in \\
		mmWave	MU-Massive-MIMO-OFDM:\\
		Genetic Algorithm based Resource Allocation
		\vspace{-1ex}
		}
\thanks{This work was supported in part by Huawei Technologies Canada and in part by the Natural Sciences and Engineering Research Council of Canada.}
}

\author{\IEEEauthorblockN{Asil Koc, Farhan Bishe, Tho Le-Ngoc}
	\IEEEauthorblockA{
		Department of Electrical and Computer Engineering, McGill University, Montreal, QC, Canada \\
		Email: 
		asil.koc@mail.mcgill.ca,
		farhan.bishe@mail.mcgill.ca,
		tho.le-ngoc@mcgill.ca
		\vspace{-2ex}
	}
}

\maketitle

\begin{abstract}
	This paper develops a new genetic algorithm based resource allocation (GA-RA) technique {for energy-efficient throughout maximization in multi-user massive multiple-input multiple-output (MU-mMIMO) systems} using orthogonal frequency division multiplexing (OFDM) based transmission.
	{We employ a hybrid precoding (HP) architecture with three stages:} (i) radio frequency (RF) beamformer, (ii) baseband (BB) precoder, (iii) resource allocation (RA) block.  
	First, a single RF beamformer block is built for all subcarriers via the slow time-varying angle-of-departure (AoD) information.
		{For enhancing the energy efficiency, the RF beamformer aims to reduce the hardware cost/complexity and total power consumption via a low number of RF chains.
		Afterwards,} the reduced-size effective channel state information (CSI) is utilized in the design of a distinct BB precoder and RA block for each subcarrier.
	The BB precoder is developed via regularized zero-forcing technique. 
		{Finally, the RA block is built via the proposed GA-RA technique for throughput maximization by allocating the power and subcarrier resources.}
	The illustrative results show that the {throughput} performance in the MU-mMIMO-OFDM systems is greatly enhanced via the proposed GA-RA technique compared to {both equal RA (EQ-RA) and particle swarm optimization based RA (PSO-RA). Moreover,} the {performance gain ratio} increases with the increasing number of subcarriers, particularly for low transmission powers.
\end{abstract}

\begin{IEEEkeywords}
	Massive MIMO, 
	hybrid precoding,
	OFDM,
	genetic algorithm,
	resource allocation,
	multi-user,
	beamforming.
\end{IEEEkeywords}

\vspace{-1ex}

\section{Introduction}
\IEEEPARstart{M}{illimeter} wave (mmWave) communication has attracted considerable attention for the fifth-generation (5G) and beyond wireless communication systems by addressing bandwidth limitation issue \cite{Uwaechia2020}. 
	{
		Bandwidth demands have arisen in accordance with the intense growth in user demands in wireless networks and emerging advanced technologies requiring high-throughput data communications.
	}
Hence, the mmWave communications is an enabling technology by providing a considerable large bandwidth (i.e., 30 GHz - 300 GHz).
	Moreover, transmitting information on such high frequencies favorizes implementing large-dimensional antenna arrays in relatively smaller dimensions. Nevertheless, making a shift to mmWave communications leads to some challenging issues, notably higher atmospheric absorption causing severe path loss due to its short wavelength\cite{Uwaechia2020}. To alleviate the path loss effect in mmWave communication, it is required to form high-gain directional beams, which can be done by implementing large antenna arrays, introducing the massive multiple-input multiple-output (mMIMO) technology\cite{5G_Mas_MIMO_mmWave_5}.

Hybrid precoding (HP) is a promising  technique for the  multi-user mMIMO (MU-mMIMO) systems \cite{Mass_MIMO_Hyb_Survey}. 
{The two-stage HP architecture interconnects the radio frequency (RF) and baseband (BB) stages via a limited number of power-hungry RF chains \cite{ASIL_FD_SIC,ASIL_ABHP_Access,
		ASIL_EE_2D_OJ_COMS,ASIL_PSO_PA_WCNC,ASIL_Subconnected_GC,ANALOG_BF_Heath}.
	Thus, HP remarkably enhances the energy-efficiency compared to the single-stage fully digital precoding (FDP).}
	On the other hand, the wideband signal transmission is considered to achieve extremely high data rates, especially in the mmWave communications. 
		To address the frequency selectivity issue, the orthogonal frequency division multiplexing (OFDM) technique is widely considered for the wideband signal transmission \cite{SU_mMIMO_OFDM}. 
	In \cite{SU_mMIMO_OFDM,mMIMO_OFDM_WSR,mMIMO_OFDM_Partially_Connected}, 
OFDM-based HP scheme is investigated for mMIMO systems, where one RF beamformer is designed for all subcarriers, while the number of BB precoders should be the same as the number of subcarriers.


{ The recent advances in artificial intelligence (AI) enable the development of advanced communication systems \cite{6G_AI}. 
	In the field of AI,}
	the genetic algorithm (GA) is a evolutionary population-based optimization technique {to solve the complex real-world problems under given constraints} \cite{kramer2017genetic,mirjalili2019evolutionary}.

In this {paper}, we propose a novel genetic algorithm based resource allocation (GA-RA) technique {for energy-efficient throughput maximization in} the MU-mMIMO-OFDM system. 
Also, the angular-based hybrid precoding (AB-HP) technique proposed in \cite{ASIL_ABHP_Access} is adopted for the OFDM-based transmission. Thus, the system model is divided into three stages: (i) RF beamformer, (ii) BB precoder, (iii) resource allocation (RA). 
	First, the RF beamformer is designed via the slow time-varying angle-of-departure (AoD) information. 
Next, the BB precoder is built for each subcarrier using the reduced-size effective channel state information (CSI) for the corresponding subcarrier.	
	Finally, the proposed GA-RA technique is applied {for the throughput maximization} by efficiently allocating the power and subcarrier resources among the users.
The numerical results present that the proposed GA-RA considerably improves the {throughput (i.e., sum-rate)\footnote{We use the terms of throughput and sum-rate interchangeably.}} performance compared to the equal RA (EQ-RA) scheme. Also, when the number of subcarriers increases, the {performance} gain improves, especially for the low transmit powers.

The rest of this paper is organized as follows. The system model is introduced in Section \ref{SEC_SYS}. The AB-HP scheme is expressed in Section \ref{SEC_HP}. Then, Section \ref{SEC_PF} provides the RA problem formulation for the {throughput} maximization. The proposed GA-RA technique is introduced in Section \ref{SEC_RA}. After presenting the illustrative results in Section \ref{SEC_RESULTS}, the paper is concluded in Section \ref{SEC_CONC}.

\section{System Model}\label{SEC_SYS}

\begin{figure}[t]
	\centering
	\includegraphics[width = \columnwidth]{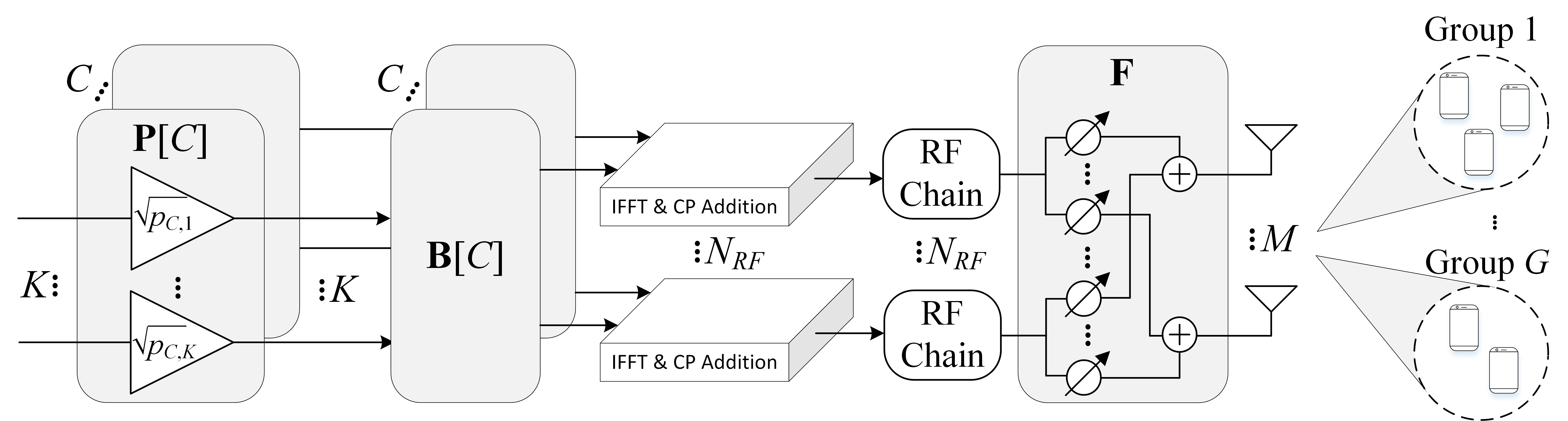}
	\caption{Hybrid precoding and genetic algorithm based resource allocation (GA-RA) in MU-mMIMO-OFDM systems.}
	\label{fig_System_Model}
\end{figure}

We consider an OFDM-based downlink transmission via $C$ subcarriers for the MU-mMIMO systems as shown in Fig. \ref{fig_System_Model}. 
	Here, $K$ single-antenna users located in $G$ groups are served by a BS having a uniform rectangular array (URA) with $M=M_x\times M_y$ antennas\footnote{
			Based on the URA structure, $M_x$ and $M_y$ denote the number of antennas along $x$-axis and $y$-axis, respectively. 
				There are two main reasons for utilizing the URA structure: (i) it packs a larger number of antennas in a two-dimensional (2D) grid under the physical-limited space requirements in practical applications, (ii) it enables three-dimensional (3D) beamforming by employing both azimuth and elevation domains \cite{5G_Mas_MIMO_mmWave_5,ASIL_ABHP_Access,ASIL_PSO_PA_WCNC,mMIMO_OFDM_Partially_Connected,ASIL_Xiaoyi_DL_CE,ASIL_EE_2D_OJ_COMS}. 
	}.
The BS utilizes $N_{RF}$ RF chains for interconnecting the BB-stage and RF-stage.
According to the multi-carrier transmission, the data signal ${\bf{d}}\hspace{-0.5ex}\left[i\right]$ for each subcarrier $i=1,\cdots,C$ is respectively passed through a power allocation block ${\bf{P}}\hspace{-0.5ex}\left[i\right] = \textrm{diag}\left(\sqrt{p_{1,i}},\cdots,\sqrt{p_{K,i}}\right)\in\mathbb{R}^{K\times K}$ and a digital BB precoder ${\bf B}\hspace{-0.5ex}\left[i\right] = \left[{\bf b}_{1,i},\cdots,{\bf b}_{K,i}\right] \in\mathbb{C}^{N_{RF} \times K}$, where $p_{k,i}$ is the allocated power for the $k^{th}$ user at subcarrier $i$.
	After applying the inverse fast Fourier transforms (IFFT) and adding the cyclic prefixes (CP), the analog RF beamformer ${\bf F}\in\mathbb{C}^{M\times N_{RF}}$ developed via low-cost phase-shifters is employed, which is identical for all subcarriers\footnote{Since the analog RF beamformer is designed via the AoD information, it is reasonable to assume that each subcarrier experiences a similar AoD support (i.e., mean azimuth/elevation AoD and their spread).}  \cite{SU_mMIMO_OFDM,mMIMO_OFDM_WSR,mMIMO_OFDM_Partially_Connected}.
Then, the transmitted data signal at subcarrier $i$ is written as:
\begin{equation}\label{eq_data_s}
	{\bf s}\hspace{-0.5ex}\left[i\right]
	=
	{\bf FB}\hspace{-0.5ex}\left[i\right]\hspace{-0.5ex}
	{\bf P}\hspace{-0.5ex}\left[i\right]\hspace{-0.5ex}
	{\bf d}\hspace{-0.5ex}\left[i\right]\in\mathbb{C}^M.
\end{equation}
	Considering the URA structure \cite{ASIL_ABHP_Access} and the 3D geometry-based mmWave channel model \cite{ChannelModels}, the channel vector for the $k^{th}$ user at subcarrier $i=1,\cdots,C$ is defined as:
\begin{equation}\label{eq_h_k}
	{\bf{h}}_k^T\hspace{-0.5ex}\left[i\right]
	\hspace{-0.5ex}=\hspace{-1ex}\sum_{l=1}^{L}\hspace{-0.5ex} \tau^{-\eta}_{k_l,i}z_{k_l,i}{\bm{\phi }}^T\hspace{-0.5ex}\big( \hspace{-0.25ex}{{\gamma _{x,k_l,i}},\hspace{-0.25ex}{\gamma _{y,k_l,i}}}\hspace{-0.25ex}\big) 
	\hspace{-0.25ex}e^{-j\frac{2\pi l i}{C}}\hspace{-0.5ex}=\hspace{-0.25ex} {\bf{z}}^T_k\hspace{-0.5ex}\left[i\right]\hspace{-0.5ex}{{\bf{\Phi }}}_k\hspace{-0.5ex}\left[i\right]\hspace{-0.5ex},
\end{equation}
where 
$L$ is the number of paths,
$\tau_{k_l,i}$ and $z_{k_l,i}\sim\hspace{-0.25ex}\mathcal{CN}\big(0,\frac{1}{L}\big)$ are respectively the distance and complex path gain of $l^{th}$ path at subcarrier $i$,
$\eta$ is the path loss exponent,
${\bm{\phi }}\big( \cdot,\cdot\big)\in \mathbb{C}^M$ is the phase response vector,
$\gamma_{x,k_l,i}\hspace{-0.5ex}=\hspace{-0.25ex}\sin\left(\theta_{k_l,i}\right)\cos\left(\psi_{k_l,i}\right)$
and
$\gamma_{y,k_l,i}\hspace{-0.5ex}=\hspace{-0.25ex}\sin\left(\theta_{k_l,i}\right)\sin\left(\psi_{k_l,i}\right)$
are the coefficients reflecting the elevation AoD (EAoD) and azimuth AoD (AAoD) for the corresponding path.
Here,
$\theta_{k_l,i}\in\big[\theta_k - \delta_k^\theta, \theta_k + \delta_k^\theta\big]$ is the EAoD with mean $\theta_k$ and spread $\delta_k^\theta$, 
$\psi_{k_l,i}\in\big[\psi_k - \delta_k^\psi, \psi_k + \delta_k^\psi\big]$ is the AAoD with mean $\psi_k$ and spread $\delta_k^\psi$.
	Then, the phase response vector is modeled as \cite{ASIL_ABHP_Access}:
\begin{equation}\label{eq_phase_vector}
	\begin{aligned}
		{\bm{\phi}}\hspace{-0.5ex}\left( {{\gamma_x, \gamma_y}} \right) \hspace{-0.5ex}&=\hspace{-0.75ex} \big[ {1,{e^{ -j2\pi d  {{\gamma_x }} }}, \cdots,{e^{ -j2\pi d\left( {{M_{x}} - 1} \right) {{\gamma_x}} }}} \big]^T\\
		&\otimes \hspace{-0.5ex}\big[ {1,{e^{-j2\pi d  {{\gamma_y }} }}, \cdots,{e^{-j2\pi d\left( {{M_{y}} - 1} \right) {{\gamma_y}} }}} \big]^T\in \mathbb{C}^M,
	\end{aligned}
\end{equation}
where $d$ is the antenna element spacing normalized by the wavelength.
As indicated in \eqref{eq_h_k}, the instantaneous channel vector is represented via the fast time-varying path gain vector 
${\bf z}_k = \left[\tau^{-\eta}_{k_1}z_{k_1},\cdots,\tau^{-\eta}_{k_L}z_{k_L}\right]^T\in\mathbb{C}^L$
and
slow time-varying phase response matrix 
	${{\bf{\Phi }}}_k\in\mathbb{C}^{L\times M}$ as a function of AoD information.
		
By using \eqref{eq_data_s} and \eqref{eq_h_k}, the received signal at the $k^{th}$ user at subcarrier $i$ is given by:
\begin{equation}
	\begin{aligned}
		{r_{\hspace{-0.1ex}k,\hspace{-0.1ex}i}} \hspace{-0.5ex}&=\hspace{-0.5ex} {\bf{h}}_{k}^T\hspace{-0.25ex}\left[i\right]{\bf{s}}\hspace{-0.25ex}\left[i\right] + {w_{k,i}} = {\bf{h}}_{k}^T\hspace{-0.25ex}\left[i\right]{\bf{FB}}\hspace{-0.25ex}\left[i\right]{\bf P}\hspace{-0.25ex}\left[i\right]{\bf d}\hspace{-0.25ex}\left[i\right] + {w_{k,i}} \\
		&=\hspace{-0.75ex} \underbrace {\sqrt {{p_{{k,i}}}} {\bf{h}}_{{k}}^T\hspace{-0.75ex}\left[i\right]\hspace{-0.5ex}{\bf{F}}{{\bf{b}}_{{k,i}}}{d_{{k,i}}}}_{\textrm{Desired Signal}} 
		\hspace{-0.25ex}+\hspace{-0.5ex}
		\underbrace {\sum_{t \ne k}^{{K}}\hspace{-0.75ex} {\sqrt {{p_{{t,i}}}} {\bf{h}}_{{k}}^T\hspace{-0.75ex}\left[i\right]\hspace{-0.5ex}{\bf{F}}{{\bf{b}}_{{t,i}}}{d_{{t,i}}}} }_{\textrm{Inter-User Interference}}
		\hspace{-0.25ex}+\hspace{-0.25ex}
		\underbrace {w_{k,i}}_{\textrm{Noise}}\hspace{-0.25ex},
	\end{aligned}
\end{equation}
where
	$w_{k,i}\sim\mathcal{CN}\big(0,\sigma^2\big)$ is the complex circularly symmetric Gaussian noise.
Then, we derive the instantaneous sum-rate (i.e., {throughput}) expression at subcarrier $i$ as follows:
\begin{equation}\label{eq_Rate_i}
	R_{\textrm{sum},i}\hspace{-.6ex}\left(
	{\bf F}\hspace{-.25ex},\hspace{-.35ex}{\bf B}\hspace{-0.65ex}\left[i\right]\hspace{-.5ex},
	\hspace{-.35ex}{\bf P}\hspace{-0.65ex}\left[i\right]\right)
	\hspace{-0.75ex}=\hspace{-0.75ex}
	\sum_{k=1}^{K}\hspace{-.35ex}\log_2\hspace{-.85ex}\left(\hspace{-.85ex}1\hspace{-.5ex}+\hspace{-.5ex}\frac{{{p_{{k,i}}}} |{\bf{h}}_{{k}}^T\left[i\right]{\bf{F}}{{\bf{b}}_{{k,i}}}|^2}
	{
		\sum\limits_{t\ne k}^{K}\hspace{-0.5ex}{{p_{{t,i}}}} \hspace{-0.25ex}|{\bf{h}}_{{k}}^T\hspace{-.5ex}\left[i\right]\hspace{-0.5ex}{\bf{F}}{{\bf{b}}_{{t,i}}}\hspace{-0.25ex}|^{\hspace{-0.1ex}2}
		\hspace{-0.65ex}+\hspace{-0.5ex}
		\sigma^2
	}\hspace{-1ex}\right)\hspace{-1ex}.
\end{equation}
{The overall sum-rate across all $C$ subcarriers is calculated as 
	$R_{\textrm{sum}}=\sum_{i=1}^{C}\mathbb{E}\left\{
	R_{\textrm{sum},i}\left({\bf F}\hspace{-.25ex},\hspace{-.35ex}{\bf B}\hspace{-0.65ex}\left[i\right]\hspace{-.5ex},
	\hspace{-.35ex}{\bf P}\hspace{-0.65ex}\left[i\right]\right)
	\right\}$
in the unit of [bps/Hz].
	Similarly, the average sum-rate per subcarrier is obtained as $\frac{R_{\textrm{sum}}}{C}$
in the unit of [bps/Hz/subcarrier]. 
Hence, we formulate the throughput maximization problem as follows:}
\begin{equation}\label{eq_OPT_1}
	\begin{aligned}
		\max_{\left\{
			{\bf F},
			{\bf B}\hspace{-0.2ex}\left[i\right],
			{\bf P}\hspace{-0.2ex}\left[i\right]
		\right\}} \hspace{-4ex}&\hspace{4ex} 
		R_{\textrm{sum},i}\left(
			{\bf F},
			{\bf B}\hspace{-0.5ex}\left[i\right],
			{\bf P}\hspace{-0.5ex}\left[i\right]
		\right)\\
		\textrm{s.t.}~~
		& \mathbb{E}\left\{\hspace{-0.5ex}\left\|{\bf{s}}\hspace{-0.5ex}\left[i\right]\right\|^2_2\hspace{-0.5ex}\right\} \hspace{-0.5ex}=\hspace{-0.5ex} \sum_{k = 1}^K \hspace{-0.25ex}{p_{k,i}}\hspace{-0.5ex}\left\|{\bf{F}}{{\bf{b}}_{k,i}}\right\|_2^2 \hspace{-0.5ex}\le\hspace{-0.5ex} \frac{P_T}{C}, 
		\hspace{1ex}p_{k,i}\hspace{-0.5ex}\ge \hspace{-0.5ex}0, \forall k,\\
		& |\left[{\bf{F}}\right]_{m,n}|=\frac{1}{\sqrt{M}},\forall m,n,\\
	\end{aligned}
\end{equation}
where
the constraints indicate the transmit power constraint per subcarrier \cite{SU_mMIMO_OFDM} (i.e., $\frac{P_T}{C}$) and the constant modulus (CM) constraint due to the utilization of phase-shifters at the RF-stage.
However, it is a non-convex optimization due to the allocated powers entangled with each other (please see \eqref{eq_Rate_i}) and the CM constraint at the RF beamformer. 
	Thus, we first design the RF beamformer and the BB precoder via AB-HP in Section \ref{SEC_HP}, then we propose the GA-RA technique in Section \ref{SEC_RA} for optimizing the allocated powers across all subcarriers.


\section{Hybrid Precoding}\label{SEC_HP}

In this section, we develop the angular-based hybrid precoding (AB-HP) architecture for MU-mMIMO-OFDM systems to reduce the number of RF chains, suppress the inter-user interference and lower the CSI overhead size.
	After designing the RF beamformer and BB precoder in this section, the proposed GA-RA technique is expressed in Section \ref{SEC_RA}.

\subsection{RF Beamformer}\label{SSEC_RF}

The slow-time varying AoD information\footnote{In \cite{ASIL_Xiaoyi_DL_CE}, it is shown that the AoD parameters (i.e., mean and spread) can be efficiently obtained via an offline deep learning and geospatial data-based estimation technique instead of the traditional online channel sounding.} is utilized at the RF beamformer for reducing the large CSI overhead size in MU-mMIMO-OFDM systems.
	Considering the $K$ users located in $G$ groups as shown in Fig. \ref{fig_System_Model}, the channel matrix for group $g$ at subcarrier $i$ is given by:
\begin{equation}\label{eq_H_g}
	{\bf H}_g \hspace{-0.5ex}\left[i\right] = \left[{\bf h}_1\hspace{-0.5ex}\left[i\right],\cdots,{\bf h}_{K_g}\hspace{-0.5ex}\left[i\right]\right]^T= {\bf Z}_g\hspace{-0.5ex}\left[i\right]{\bf\Phi}_g\hspace{-0.5ex}\left[i\right]\in\mathbb{C}^{K_g\times M},
\end{equation}
where 
${\bf Z}_g \hspace{-0.5ex}\left[i\right]\hspace{-0.25ex}=\hspace{-0.5ex} \big[
	{\bf z}_1\hspace{-0.5ex}\left[i\right],\cdots, 
	{\bf z}_{K_g}\hspace{-0.5ex}\left[i\right]\big]^T\hspace{-0.5ex}\in\hspace{-0.25ex}\mathbb{C}^{K_g\times L}$,
	${\bf \Phi}_g\hspace{-0.5ex}\left[i\right]\hspace{-0.5ex}\in\hspace{-0.25ex}\mathbb{C}^{L\times M}$,
$K_g$ is the number of users in group $g$ with $K=\sum_{g=1}^{G}K_g$.
Then, the full-size channel matrix at subcarrier $i$ is defined as 
	${\bf H} \hspace{-0.25ex}\left[i\right]\hspace{-0.5ex}= \hspace{-0.5ex}
	\left[
	{\bf H}_1^T\hspace{-0.25ex}\left[i\right],
	\cdots,
	{\bf H}_G^T\hspace{-0.25ex}\left[i\right]
	\right]^T
	\hspace{-0.5ex}\in\hspace{-0.25ex}\mathbb{C}^{K\times M}$.
		Here, we assume that the users in the same groups experience a similar AoDs as in \cite{ASIL_ABHP_Access,ASIL_PSO_PA_WCNC,ASIL_EE_2D_OJ_COMS,ASIL_Subconnected_GC}. 
Therefore, $G$ blocks are designed for the RF beamformer as follows:
\begin{equation}\label{eq_F}
	{\bf F}= \left[{\bf F}_1,\cdots,{\bf F}_G\right]\in \mathbb{C}^{M\times N_{RF}},
\end{equation}
where 
${\bf F}_g \hspace{-0.25ex}\in \hspace{-0.25ex}\mathbb{C} ^{M\times N_{RF,g}}$ is the RF beamformer for group $g$
with 
$N_{RF}\hspace{-0.45ex}=\hspace{-0.55ex}\sum\nolimits_{g=1}^{G}\hspace{-0.25ex}N_{RF,g}$.
	By using \eqref{eq_H_g} and \eqref{eq_F}, the effective channel matrix seen from the BB-stage is obtained as:
\begin{equation}\label{eq_H_eff}
	\bm{\mathcal{H}}\hspace{-0.5ex}\left[i\right]
	\hspace{-0.75ex}=\hspace{-0.5ex}
	{\bf H}\hspace{-0.5ex}\left[i\right]\hspace{-0.5ex}{\bf F} \hspace{-0.5ex}= \hspace{-1ex}
	\left[ \hspace{-1.5ex}{\begin{array}{*{20}{c}}
			{{{\bf{H}}_1}\hspace{-0.5ex}\left[i\right]\hspace{-0.5ex}{{\bf{F}}_1}}&\hspace{-2ex}{{{\bf{H}}_1}\hspace{-0.5ex}\left[i\right]\hspace{-0.5ex}{{\bf{F}}_2}}&\hspace{-2ex}{\cdots}&\hspace{-2ex}{{{\bf{H}}_1}\hspace{-0.5ex}\left[i\right]\hspace{-0.5ex}{{\bf{F}}_G}}\\
			{{{\bf{H}}_2}\hspace{-0.5ex}\left[i\right]\hspace{-0.5ex}{{\bf{F}}_1}}&\hspace{-2ex}{{{\bf{H}}_2}\hspace{-0.5ex}\left[i\right]\hspace{-0.5ex}{{\bf{F}}_2}}&\hspace{-2ex}{\cdots}&\hspace{-2ex}{{{\bf{H}}_2}\hspace{-0.5ex}\left[i\right]\hspace{-0.5ex}{{\bf{F}}_G}}\\
			{\vdots}&\hspace{-2ex}{\vdots}&\hspace{-2ex}{\ddots}&\hspace{-2ex}{\vdots}\\
			{{{\bf{H}}_G}\hspace{-0.5ex}\left[i\right]\hspace{-0.5ex}{{\bf{F}}_1}}&\hspace{-2ex}{{{\bf{H}}_G}\hspace{-0.5ex}\left[i\right]\hspace{-0.5ex}{{\bf{F}}_2}}&\hspace{-2ex}{\cdots}&\hspace{-2ex}{{{\bf{H}}_G}\hspace{-0.5ex}\left[i\right]\hspace{-0.5ex}{{\bf{F}}_G}}\\
	\end{array}} \hspace{-1.5ex}\right]\hspace{-1.25ex}\in\hspace{-0.5ex}\mathbb{C}^{K\hspace{-0.25ex}\times\hspace{-0.25ex} N_{RF}}\hspace{-0.5ex},
\end{equation}
where the diagonal block matrix ${\bf H}_g\hspace{-0.5ex}\left[i\right]\hspace{-0.5ex}{\bf F}_g={\bf Z}_g\hspace{-0.5ex}\left[i\right]\hspace{-0.5ex}{\bf \Phi}_g\hspace{-0.5ex}\left[i\right]\hspace{-0.5ex}{\bf F}_g\in\mathbb{C}^{K_g\times N_{RF,g}}$ is the effective channel matrix and the off-diagonal block matrix ${\bf H}_t\hspace{-0.5ex}\left[i\right]\hspace{-0.5ex}{\bf F}_g\hspace{-0.5ex}=\hspace{-0.5ex}{\bf Z}_t\hspace{-0.5ex}\left[i\right]\hspace{-0.5ex}{\bf \Phi}_t\hspace{-0.5ex}\left[i\right]\hspace{-0.5ex}{\bf F}_g\in\mathbb{C}^{K_t\times N_{RF,g}}$ is the effective interference channel matrix, $\forall t\hspace{-0.5ex}\neq\hspace{-0.5ex} g$.
Therefore, the RF beamformer design targets for achieving the following two objectives: (i) maximizing the beamforming gain in the desired direction 
	(i.e., $\textrm{Span}\left({\bf F}_g\right) \subset  \textrm{Span}\left({\bf \Phi}_g\hspace{-0.5ex}\left[i\right]\hspace{-0.25ex}\right)$),
(ii) successfully suppress the interference among user groups 
	(i.e., $\textrm{Span}\left({\bf F}_g\right) \subset  \cup_{t\ne g}\textrm{Null}\left({\bf \Phi}_t\hspace{-0.5ex}\left[i\right]\hspace{-0.25ex}\right)$).
As proven in \cite{ASIL_ABHP_Access}, both objectives are accomplished by building the RF beamformer ${\bf F}_g$ via the steering vector 
	${\bf e}\left(\gamma_x,\gamma_y\right)\hspace{-0.25ex}=\hspace{-0.25ex}\frac{1}{\sqrt{M}}{\bm{\phi}}^*\hspace{-0.25ex}\left( {{\gamma_x, \gamma_y}} \right)\hspace{-.25ex}\in\hspace{-0.25ex}\mathbb{C}^M$ with $\left(\gamma_x,\gamma_y\right)$ angle-pairs covering the AoD support of desired user group and excluding the AoD supports of the other user groups
	(please see \eqref{eq_phase_vector} for ${\bm{\phi}}\hspace{-0.25ex}\left( {{\gamma_x, \gamma_y}} \right)$).
For covering the complete 3D elevation and azimuth angular space with minimum number of angle-pairs, $M$ orthogonal quantized angle-pairs are defined as
	${{\lambda^x_{u}} \hspace{-0.5ex}=\hspace{-0.5ex} -1 + \frac{2u-1}{{{M_{x}}}}}$ 
	for 
	$u = 1, \cdots,{M_{x}}$
	and 
	${{\lambda^y_{n}} = -1 + \frac{2n-1}{{{M_{y}}}}}$ 
	for
	$n = 1, \cdots,{M_{y}}$.
Considering that $N_{RF,g}$ quantized angle-pairs covers the AoD support of user group $g$ \cite[eq. (13)]{ASIL_ABHP_Access}, we build the RF beamformer for group $g$ as follows:
\begin{equation}\label{eq_F_g}
	{\bf F}_g
	\hspace{-0.5ex}=\hspace{-0.5ex}
	\big[\hspace{-0.25ex}{\bf e}\big(\hspace{-.25ex}\lambda_{u_1}^x,\hspace{-0.25ex}\lambda_{n_1}^y\hspace{-0.25ex}\big),\hspace{-0.25ex}\cdots\hspace{-0.25ex},
	\hspace{-.25ex}{\bf e}\big(\hspace{-.25ex}\lambda_{u_{N_{RF,g}}}^x\hspace{-.5ex},\hspace{-.25ex}\lambda_{n_{N_{RF,g}}}^y\hspace{-.25ex}\big)\hspace{-.25ex}\big]\hspace{-.5ex}\in\hspace{-.5ex}\mathbb{C}^{M\hspace{-.25ex}\times N_{RF,g}}\hspace{-.25ex}.
\end{equation}
Finally, the complete RF precoder ${\bf F}$ satisfying the CM constraint given in \eqref{eq_OPT_1} is derived by  substituting \eqref{eq_F_g} into \eqref{eq_F}.
	It is worthwhile to mention that the RF beamformer is a unitary matrix (i.e., ${\bf F}^H {\bf F} = {\bf I}_{N_{RF}}$).
\subsection{BB Precoder}\label{SSEC_BB}
As seen in Fig. \ref{fig_System_Model}, we develop $C$ distinct BB precoders for each subcarriers. 
		{Here, the main objective is to further suppress the residual inter-user interference.}
	By utilizing the reduced-size effective channel matrix $\bm{\mathcal{H}}\hspace{-0.5ex}\left[i\right]$, the regularized zero-forcing (RZF) technique is applied for each subcarrier.
Thus, the BB precoder at subcarrier $i$ is defined as:
\begin{equation}\label{eq_BB}
	{\bf B}\hspace{-0.5ex}\left[i\right] = \big[\hspace{-0.5ex}\left(\bm{\mathcal{H}}\hspace{-0.5ex}\left[i\right]\right)^H\bm{\mathcal{H}}\hspace{-0.5ex}\left[i\right] + K\alpha{\bf I}_K\big]^{-1}\hspace{-0.5ex}\left(\bm{\mathcal{H}}\hspace{-0.5ex}\left[i\right]\right)^H\in\mathbb{C}^{N_{RF}\times K},
\end{equation}
where $\alpha=\frac{\sigma^2}{P_T}$ is the regularization parameter \cite{ASIL_ABHP_Access}.


\section{Problem Formulation}\label{SEC_PF}
After deriving the closed-form solutions for RF beamformer ${\bf F}$ and BB precoder ${\bf B}\hspace{-0.5ex}\left[i\right]$ for all subcarriers $i=1,\cdots,C$, the {throughput maximization} problem given in \eqref{eq_OPT_1} turns into a resource allocation (RA) problem (i.e., allocating power and subcarrier resources among the users).
	Thus, for a given ${\bf F}$ and ${\bf B}\left[i\right]$, we formulate the RA optimization problem as follows:
\begin{equation}\label{eqn: Optimization}
	\begin{aligned}
		\max_{
			{\bf P}\hspace{-0.2ex}\left[i\right]
		} ~& 
		R_{\textrm{sum},i}\left(
		{\bf F},
		{\bf B}\hspace{-0.5ex}\left[i\right],
		{\bf P}\hspace{-0.5ex}\left[i\right]
		\right)\\
		\textrm{s.t.}~~~
		& \mathbb{E}\left\{\hspace{-0.5ex}\left\|{\bf{s}}\hspace{-0.5ex}\left[i\right]\right\|^2_2\hspace{-0.5ex}\right\} \hspace{-0.5ex}=\hspace{-0.5ex} \sum_{k = 1}^K {{p_{k,i}}{\bf{b}}_{k,i}^H{{\bf{b}}_{k,i}}} \hspace{-0.5ex}\le\hspace{-0.5ex} \frac{P_T}{C},~~ p_{k,i}\hspace{-0.5ex}\ge\hspace{-0.5ex}0, \forall k,
	\end{aligned}
\end{equation}
where 
	$R_{\textrm{sum},i}\left(
	{\bf F},
	{\bf B}\hspace{-0.5ex}\left[i\right],
	{\bf P}\hspace{-0.5ex}\left[i\right]
	\right)$
is the sum-rate at subcarrier $i$ defined in \eqref{eq_Rate_i}.
	However, it is still a non-convex optimization problem due to the allocated power $p_{k,i}$ interchangeably located in the numerator and denominator in \eqref{eq_Rate_i}.
Thus, the traditional optimization algorithms may not be utilized to solve the RA problem.

\section{Genetic Algorithm based Resource Allocation}\label{SEC_RA}
Genetic algorithm (GA) is one of the most well-known nature-inspired evolutionary optimization algorithms \cite{mirjalili2019evolutionary}. It can address the shortcomings of traditional optimization algorithms, which are not able to find the optimal solution for non-convex problems. According to the GA, each solution of the problem is considered a chromosome, where the genes on a chromosome are defined as the problem variables. The chromosomes/genes evolve through generations via three main steps: (i) selection, (ii) crossover (iii) mutation \cite{mirjalili2019evolutionary,haupt2004practical}. 

	We here propose a new genetic algorithm based resource allocation (GA-RA) technique for MU-mMIMO-OFDM systems. 
	As the RA optimization problem defined in \eqref{eqn: Optimization} aims to find the optimal allocated powers for $K$ users at subcarrier $i$, it lies on $K$-dimensional search space. Thus, each chromosome contains $K$ genes. 
Since we deal with the continuous values, the continuous GA is adopted instead of the binary GA. 
Consider the $n^{th}$ chromosome representing the allocated powers of the $q^{th}$ generation at subcarrier $i$ as follows:\vspace{0.5ex}
\begin{equation}\label{ResourceAllocation_Block}
	{\bf P}_n^{(q)}[i] = \textrm{diag}\left(\sqrt{p_{1,i,n}^{(q)}},\cdots, \sqrt{p_{K,i,n}^{(q)}}\right) \in\mathbb{R}^{K\times K},\vspace{0.5ex}
\end{equation}
where $p_{k,i,n}^{\left(q\right)}$ is the $k^{th}$ user power at subcarrier $i$. For satisfying the total transmitted power constraint for each subcarrier given in \eqref{eqn: Optimization}, each chromosome is normalized as:\vspace{0.5ex}
\begin{equation}\label{Normalized_CM1}
	\hat{\bf{P}}_{n}^{(q)}\hspace{-.25ex}[i] 
	\hspace{-0.6ex}=\hspace{-0.6ex}
	\frac{{\bf{P}}_n^{(q)}\hspace{-.25ex}[i]}{\varepsilon_n^{(q)}\hspace{-.25ex}[i]}
	\hspace{-0.5ex}=\hspace{-0.25ex}
	\textrm{diag}\hspace{-0.5ex}\left(\hspace{-1ex}\sqrt{\hat{p}_{1,i,n}^{(q)}},\hspace{-0.25ex}\cdots\hspace{-0.25ex},\sqrt{\hat{p}_{K,i,n}^{(q)}}\right) \hspace{-0.5ex}\in\hspace{-0.25ex}\mathbb{R}^{K\times K}\hspace{-0.25ex},\hspace{-0.25ex}\vspace{0.5ex}
\end{equation}
where 
	$\varepsilon_{n}^{(q)}[i] = \sqrt{\frac{P_T/C}
		{
			\sum_{k = 1}^K {{\hat{p}_{k,i,n}^{(q)}}{\bf{b}}_{k,i}^H{{\bf{b}}_{k,i}}}
		}}$
and $\hat{p}_{k,i,n}^{(q)} \in [0,1]$. The fitness function corresponding to the $n^{th}$ chromosome at  subcarrier $i$ is calculated via
	$R_{\textrm{sum},i}\big(
	{\bf F},
	{\bf B}\hspace{-0.5ex}\left[i\right],
	{\bf P}_n^{\left(q\right)}\hspace{-0.5ex}\left[i\right]
	\big)$
defined in \eqref{eq_Rate_i}. In the proposed GA-RA, the population size is defined as $N_p$, which is kept as the same through $Q$ generations. The first generation is initialized randomly, where each allocated power is uniformly distributed as $\hat{p}_{k,i,n}^{\left(1\right)} \sim U(0,\kappa)$ with $\kappa\in\left(0,1\right]$. Afterwards, the top $\rho_{\textrm{mating}}$ percent of the population is selected to form the mating pool, and the rest are discarded to free up space for new offspring. In order to generate new chromosomes, we need to choose parents from the mating pool. 
As new offspring inherits the traits from their parents, if only the best ones are selected the algorithm may be trapped in a local optimum point. To prevent this issue, it is needed to choose parents on a random basis while preserving our desire to choose highly qualified parents\cite{mirjalili2019evolutionary}.
	Thus, we employ the tournament selection method \cite{miller1995genetic}. 
For selecting each parent, $\rho_{\textrm{sub}}$ percent of the mating pool is randomly chosen as a subset. Then, based on the fitness value, the best chromosome of the subset is selected as a parent.
Each pair of parents produces a new pair of  offspring through the crossover. In the proposed approach, we apply a mixture of two crossover methods: (i) linear crossover\cite{WRIGHT1991205}, (ii) uniform crossover\cite{haupt2004practical}. 
	According to the a given pair of parents as 
		$\hat{\bf{P}}_{n_1}^{\left(q\right)}[i]$
	and
		$\hat{\bf{P}}_{n_2}^{\left(q\right)}[i]$
	with the randomly chosen $n_1$ and $n_2$ indices via the tournament selection method,
the linear crossover finds $\hat{\bf{P}}_{o_1}^{\left(q\right)}[i]$ and $\hat{\bf{P}}_{o_2}^{\left(q\right)}[i]$ as the new offspring as follows:\vspace{0.5ex}
\begin{equation}\label{eqn: offspring}
	\begin{aligned}
	&\hspace{-0.25ex}\hat{\bf{P}}_{o_1}^{{\left(q\right)}}\hspace{-0.25ex}[i,k]
	\hspace{-0.5ex}=\hspace{-0.75ex}
	\begin{cases}
		\hspace{-0.5ex}\frac{
			3\hat{\bf{P}}_{n_1}^{\left(q\right)}\hspace{-.15ex}[i,k] 
			-
			\hat{\bf{P}}_{n_2}^{\left(q\right)}\hspace{-.15ex}[i,k]
		}{2},
	&
		\hspace{-1.5ex} 
		0\hspace{-0.5ex}
		\le\hspace{-0.5ex} 
		\frac{
			3\hat{\bf{P}}_{n_1}^{\left(q\right)}\hspace{-.15ex}[i,k] 
			-
			\hat{\bf{P}}_{n_2}^{\left(q\right)}\hspace{-.15ex}[i,k]
		}{2}\hspace{-0.5ex}
		\le \hspace{-0.5ex}
		1\\
		\hspace{-0.5ex}\hat{\bf{P}}_{n_1}^{\left(q\right)}[i,k],&\hspace{-1.5ex} \text{otherwise}
	\end{cases}\\
	&\hspace{-0.25ex}\hat{\bf{P}}_{o_2}^{{\left(q\right)}}\hspace{-0.25ex}[i,k]
	\hspace{-0.5ex}=\hspace{-0.75ex}
	\begin{cases}
		\hspace{-0.5ex}\frac{
			3\hat{\bf{P}}_{n_2}^{\left(q\right)}\hspace{-.15ex}[i,k] 
			-
			\hat{\bf{P}}_{n_1}^{\left(q\right)}\hspace{-.15ex}[i,k]
		}{2},
		&
		\hspace{-1.5ex} 
		0\hspace{-0.5ex}
		\le\hspace{-0.5ex} 
		\frac{
			3\hat{\bf{P}}_{n_2}^{\left(q\right)}\hspace{-.15ex}[i,k] 
			-
			\hat{\bf{P}}_{n_1}^{\left(q\right)}\hspace{-.15ex}[i,k]
		}{2}\hspace{-0.5ex}
		\le \hspace{-0.5ex}
		1\\
		\hspace{-0.5ex}\hat{\bf{P}}_{n_2}^{\left(q\right)}[i,k],&\hspace{-1.5ex} \text{otherwise}
	\end{cases}
	\end{aligned}\vspace{0.5ex}
\end{equation}
where we abuse the notation $\hat{\bf{P}}_{n_1}^{\left(q\right)}[i,k]$ to represent the $k^{th}$ diagonal entry of $\hat{\bf{P}}_{n_1}^{\left(q\right)}[i]$. 
	It is important to note that the linear crossover operation might generate infeasible solutions violating the condition of $\hat{p}_{n_K,i}^{(q)} \in [0,1]$. 
To prevent this from happening, infeasible genes are replaced by the original parent's genes. Subsequently, the uniform crossover is applied by randomly swapping the genes of $\hat{\bf{P}}_{o_1}^{\left(q\right)}[i]$ with the corresponding genes of $\hat{\bf{P}}_{o_2}^{\left(q\right)}[i]$. 
The crossover process is applied until reaching the initial population size of $N_p$. Afterwards, the top $\rho_{\textrm{elite}}$ percent of them are selected as elite members, implying that they do not undergo mutation to keep the good solutions. The rest of the population are mutated by adding a Gaussian noise distributed as $\mathcal{N}(0,10^{-4})$ to the randomly selected genes based on the mutation rate of $\rho_{\textrm{mut}}$. All mutated genes are bounded in the interval of $[0,1]$. 
	Here, we advocate the exploitative behavior of the GA-RA technique by introducing crossover, whereas the mutation encourages the exploratory behavior of the GA-RA technique for preventing it from converging to a local optimum solution.
The whole procedure is repeated through $Q$ generations. 
Finally, the best solution of the last generation is reported as the power allocation block at subcarrier $i$ as follows:
\begin{equation} \label{eqn: Optimal_Solution}
		{\bf{P}}[i] = \varepsilon_{best}^{(Q)}[i]\hat{\bm{P}}_{best}^{(Q)}[i]
\end{equation}
The summary of the proposed GA-RA technique is expressed in detail in Algorithm 1.
\begin{algorithm}[!t]
	\caption{Proposed GA-RA technique}
	\begin{algorithmic}[1]
		\renewcommand{\algorithmicrequire}{\textbf{Input:}}
		\renewcommand{\algorithmicensure}{\textbf{Output:}}
		\REQUIRE ${\bf F}$, ${\bf B}[i]$, $P_T$, $N_p$, $Q$, $\kappa$, $\rho_{\textrm{mating}}$,$\rho_{\textrm{sub}}$,$\rho_{\textrm{elite}}$, $\rho_{\textrm{mut}}$
		\ENSURE  ${\bf P}[1]$, ${\bf P}[2]$,$\cdots$, ${\bf P}[C]$
		\FOR {$i = 1:C$}
		\FOR {$n= 1:N_p$}
		\STATE Initialize $\hat{\bf{P}}_n^{\hspace{-0.25ex}\left(1\right)}\hspace{-.25ex}[i]\hspace{-0.5ex}$ with diagonal entries $\hat{p}_{k,i,n}^{\left(1\right)} \hspace{-0.5ex}\sim\hspace{-0.25ex} U\hspace{-0.25ex}(0,\kappa)$.
		\ENDFOR 
		\FOR {{$q= 1:Q$}}
		\STATE Keep top $\rho_{\textrm{mating}}$ percent of population in $q^{th}$ generation as mating pool and discard the rest.
		\FOR {$p=1:\frac{(1-\rho_{\textrm{mating}})}{2}N_p$}
		\FOR {{$j=1:2$}}
		\STATE For tournament selection, create a subset by randomly choosing $\rho_{\textrm{sub}}$ percent of mating pool.
		\STATE Select the best subset member as parent $\hat{\bf{P}}_{n_j}^{\left(q\right)}[i]$.
		\ENDFOR 
		\STATE Apply linear crossover via \eqref{eqn: offspring} to calculate new offspring $\hat{\bf{P}}_{o_1}^{\left(q\right)}[i]$ and $\hat{\bf{P}}_{o_2}^{\left(q\right)}[i]$.
		\STATE Update $\hat{\bf{P}}_{o_1}^{\left(q\right)}[i]$ and $\hat{\bf{P}}_{o_2}^{\left(q\right)}[i]$ via uniform crossover by randomly swapping their corresponding genes.
		\ENDFOR
		\STATE Include all offspring to the population.
		\STATE Select top $\rho_{\textrm{elite}}$ percent of population as elite members, mutate the rest based on mutation rate of $\rho_{\textrm{mut}}$.
		\STATE Carry \hspace{-0.1ex}elite/mutated population to \hspace{-0.1ex}$(q\hspace{-0.25ex}+\hspace{-0.25ex}1)^{th}$ generation.
		\ENDFOR  
		\STATE  Find ${\bf P}[i]$ via \eqref{eqn: Optimal_Solution}.
		\ENDFOR 
		\RETURN ${\bf P}[1]$, ${\bf P}[2]$,$\cdots$, ${\bf P}[C]$
	\end{algorithmic}
\end{algorithm}

\section{Illustrative Results}\label{SEC_RESULTS}

This section illustrates Monte-Carlo simulation results to evaluate the {sum-rate and energy-efficiency} performance of the proposed ${\textrm{GA-RA}}$ technique in MU-mMIMO-OFDM systems.
	Table \ref{tbl: Simulation_Setup} presents the simulation parameters considering the 3D microcell scenario \cite{Report_5G_UMi_UMa_Rel16}.
We have considered a BS equipped with a square URA with $M=256$ antenna elements.
	{For the proposed GA-RA technique, we assume the population size as $N_p=100$ and the number of generations as $Q=10$, unless otherwise stated.} Also, other parameters are chosen as $\kappa=\frac{1}{20}$, $\rho_{\textrm{mating}}=40\%$,$\rho_{\textrm{sub}}=10\%$, $\rho_{\textrm{elite}}=80\%$, $\rho_{\textrm{mut}}=60\%$.

{The energy-efficiency of the MU-mMIMO-OFDM systems is evaluated by taking the ratio of the overall sum-rate $R_{\textrm{sum}}$ and the total power consumption $P_{\textrm{total}}$ as \cite{ASIL_ABHP_Access,ASIL_EE_2D_OJ_COMS,ANALOG_BF_Heath}:
	\begin{equation}
		\beta
		\hspace{-0.25ex}
		=
		\hspace{-0.25ex}
		\frac{R_{\textrm{sum}}}{P_{\textrm{total}}} 
		\hspace{-0.5ex}
		=
		\hspace{-0.5ex}
		\frac{
			\sum_{i=1}^{C}\mathbb{E}\left\{
			R_{\textrm{sum},i}\left({\bf F}\hspace{-.25ex},\hspace{-.35ex}{\bf B}\hspace{-0.65ex}\left[i\right]\hspace{-.5ex},
			\hspace{-.35ex}{\bf P}\hspace{-0.65ex}\left[i\right]\right)
			\right\}
		}{
			P_T + N_{RF}P_{RF} + N_{PS}P_{PS}
		}~\hspace{-0.5ex}
		\textrm{[bps/Hz/W]},
	\end{equation}
	where $P_T$ is the transmission power,
	$P_{RF}$ ($P_{PS}$) is the power consumption per each RF chain (phase-shifter),
	$N_{RF}$ ($N_{PS}$) is the number of RF chains (phase-shifters).
	We here assume that $P_{RF}=250$ mW and $P_{PS}=1$ mW as in \cite{ANALOG_BF_Heath}.
	It is important to highlight that the FDP architecture has $N_{RF}=M=256$ RF chains and does not require any phase-shifters (i.e., $N_{PS}=0$).
On the other hand, the proposed AB-HP architecture requires $N_{RF}=36$ RF chains and $N_{PS}=N_{RF}\times M=9216$ phase-shifters to support ${M=256}$ antennas in the considered simulation setup. Thus, the number of RF chains reduces from $256$ to $36$ (i.e., $87\%$ reduction in hardware cost/complexity and CSI overhead size).

}

\begin{table}[t!]
	\caption{Simulation parameters.}
	\label{tbl: Simulation_Setup}
	\centering
	\begin{tabular}{|c|c|}
		\hline
		{\# of antennas \cite{Report_5G_UMi_UMa_Rel16}} & $M=16\times 16 = 256$      \\ \hline
		{Cell radius  \cite{Report_5G_UMi_UMa_Rel16}} &  100m\\ \hline
		{BS height \cite{Report_5G_UMi_UMa_Rel16} \big|\big. User height \cite{Report_5G_UMi_UMa_Rel16}} &  10m \big|\big. 1.5m-2.5m \\ \hline
		{User-BS horizontal distance} & 10m-90m\\ \hline
		{\# of groups} & {$G=3$}\\ \hline
		{\# of users in each group} & $K_g=\frac{K}{G}$      \\ \hline
		{Mean EAoD \big|\big. Mean AAoD} & {$\theta_g \hspace{-0.5ex}=\hspace{-0.25ex} 60^\circ$ \hspace{-0.5ex}\big|\big. \hspace{-0.5ex}$\psi_g\hspace{-0.5ex}=\hspace{-0.5ex}21^\circ \hspace{-0.5ex}+\hspace{-0.5ex} 120^\circ\hspace{-0.5ex}\left(g\hspace{-0.25ex}-\hspace{-0.25ex}1\right)$ }  \\ \hline
		{EAoD spread \big|\big. AAoD spread} & {\resizebox{0.11\hsize}{!}{$\delta_g^\theta=15^\circ$} \big|\big. \resizebox{0.11\hsize}{!}{$\delta_g^\psi=11^\circ$}}\hspace{14.1ex} \\ \hline
		{Path loss exponent} \cite{ASIL_PSO_PA_WCNC}& $\eta=3.76$      \\ \hline
		{Noise PSD} \cite{ASIL_PSO_PA_WCNC}& $-174$ dBm/Hz      \\ \hline
		{Channel bandwidth} \cite{ASIL_PSO_PA_WCNC}& $10$ kHz      \\ \hline
		{\# of paths} & $L=10$      \\ \hline
		Antenna spacing (in wavelength) & ${d\hspace{-0.25ex}=0.5}$\\ \hline
		{\# of network realizations} & $1000$      \\ \hline
	\end{tabular}
\end{table}

\begin{figure}[!t]
	\begin{subfigure}{.48\columnwidth}
		\centering
		\includegraphics[width = \textwidth]{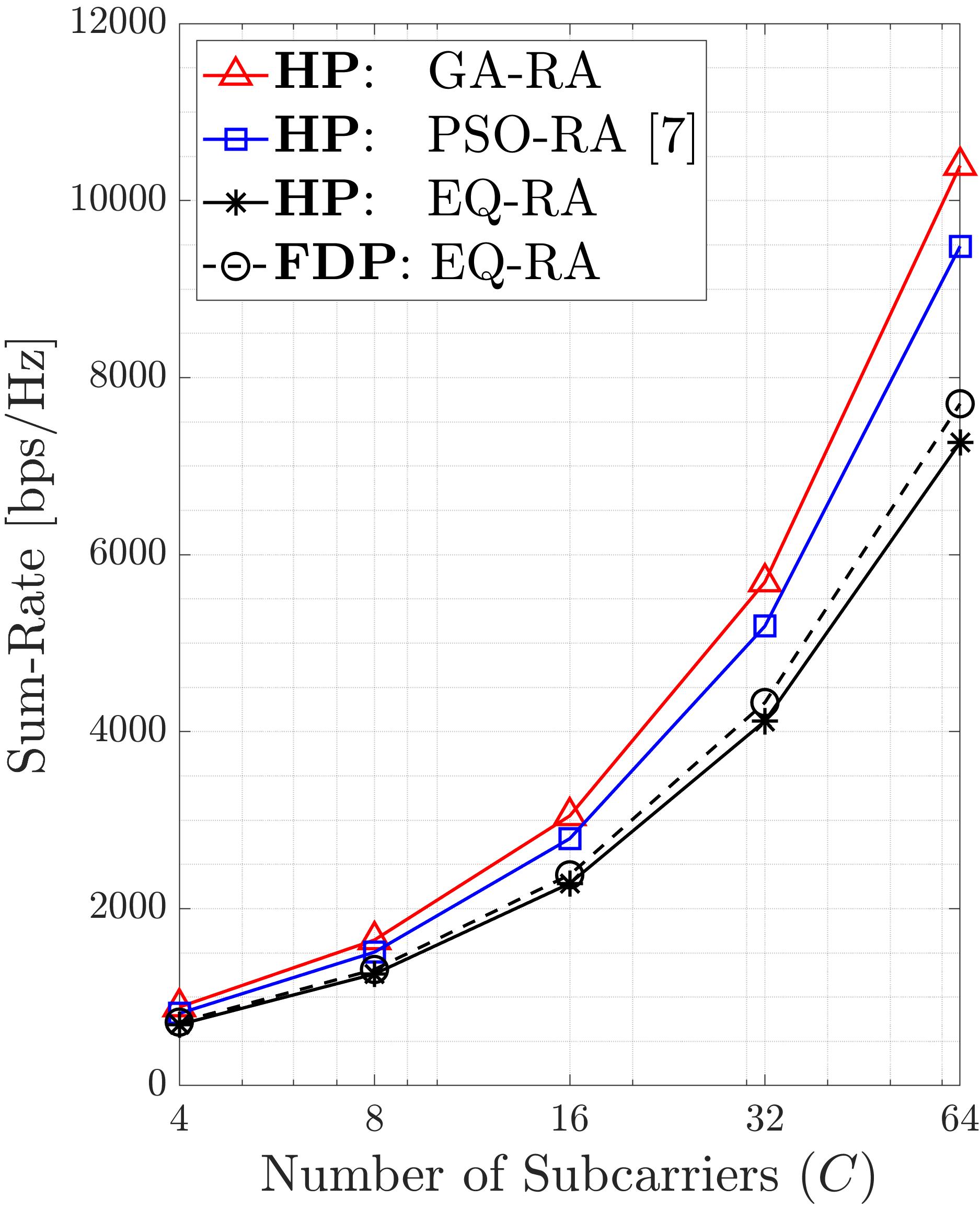}
		\caption{Sum-Rate}
	\end{subfigure}
	\hfill
	\begin{subfigure}{.49\columnwidth}
		\includegraphics[width = \textwidth]{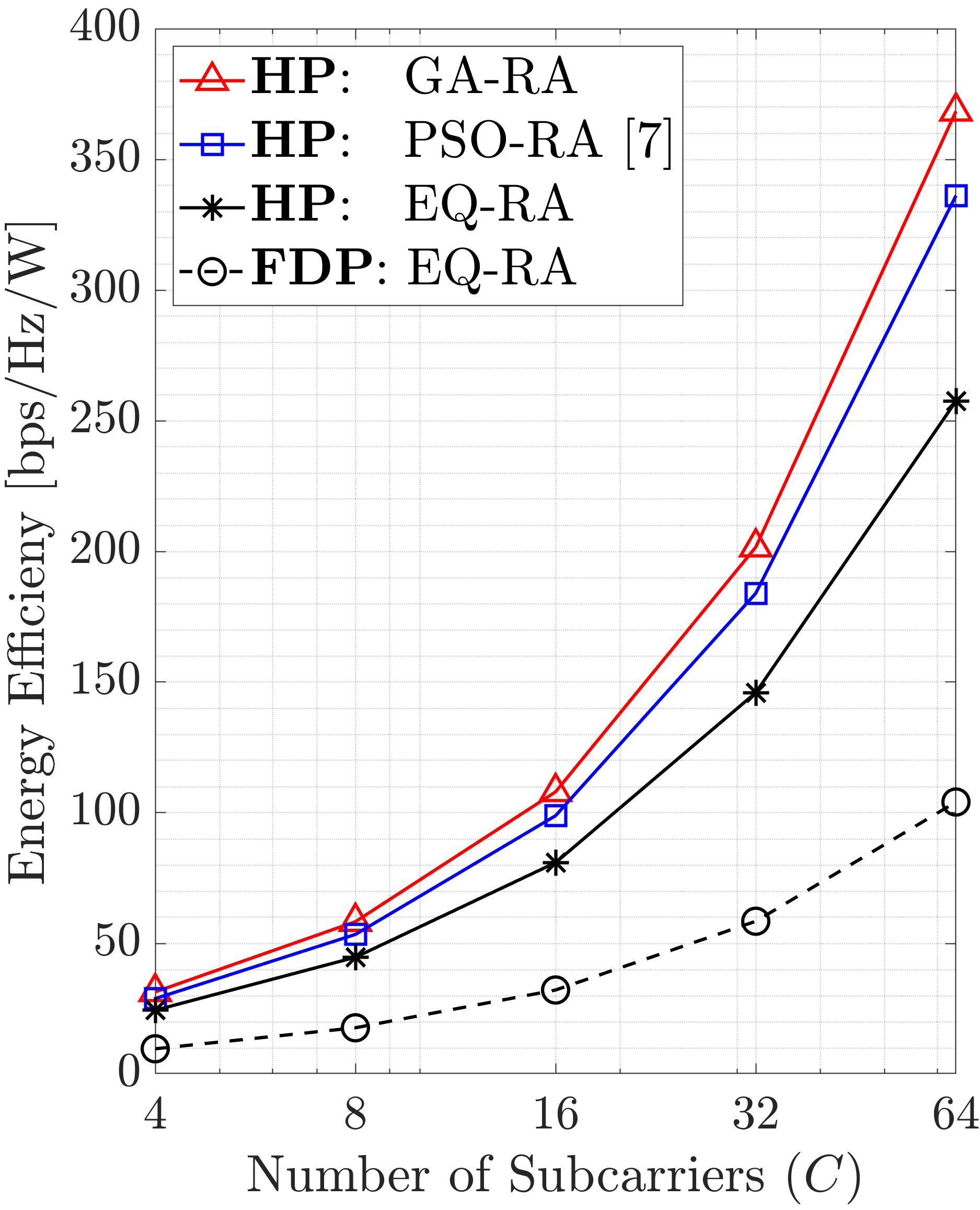}
		\caption{Energy-Efficiency}
	\end{subfigure}
	\caption{{Performance comparison among HP and FDP schemes ($P_T=40$ dBm and $K=15$ users).}}
	\vspace{-2ex}
	\label{fig: sum-rate bargraph} 
\end{figure}

{The sum-rate {and energy-efficiency} are respectively plotted versus the number of subcarriers in Fig. \ref{fig: sum-rate bargraph}(a) and Fig. \ref{fig: sum-rate bargraph}(b), where $P_T = 40$ dBm and $K = 15$ users. 
	Here, the proposed GA-RA technique is compared with {equal RA (EQ-RA)} and particle swarm optimization based RA (PSO-RA)\footnote{Even though PSO algorithm is employed for power allocation optimization over single subcarrier transmission in \cite{ASIL_PSO_PA_WCNC}, it nevertheless serves as benchmark.} \cite{ASIL_PSO_PA_WCNC}. Both GA-RA and PSO-RA techniques consider $Q=10$ generations (e.g., iterations). 
 		{Also, the power values are equally distributed among the users for all subcarriers in EQ-RA (i.e., $p_{1,1}=\cdots=p_{k,i}=\cdots=p_{K,C}$, $\forall k, i$).
	In the EQ-RA scheme, the performance comparison is investigated between FDP and HP schemes.
The numerical results in Fig. \ref{fig: sum-rate bargraph}(a) reveal that HP with EQ-RA closely achieves the sum-rate provided by its FDP counterpart.
	Additionally, Fig. \ref{fig: sum-rate bargraph}(b) illustrates that HP with EQ-RA considerably improves the energy-efficiency compared to the FDP scheme by means of reduced hardware cost/complexity.
For instance, when there are $C=16$ subcarriers, the energy-efficiency increases from $32$ bps/Hz/W to $80$ bps/Hz/W.}
In comparison to EQ-RA, the proposed GA-RA technique enhances the sum-rate and energy-efficiency by $28.6\%$, $33.5\%$, $43.1\%$ for ${C=4,16,64}$ subcarriers, respectively.
	Hence, the performance gap between GA-RA and EQ-RA keeps increasing for larger number of subcarriers.
	On the other hand, the numerical results demonstrate that the proposed GA-RA technique outperforms PSO-RA by further enhancing {both sum-rate and energy-efficiency}.}

\begin{figure}[!t]
	\begin{subfigure}{.48\columnwidth}
		\centering
		\includegraphics[width = \textwidth]{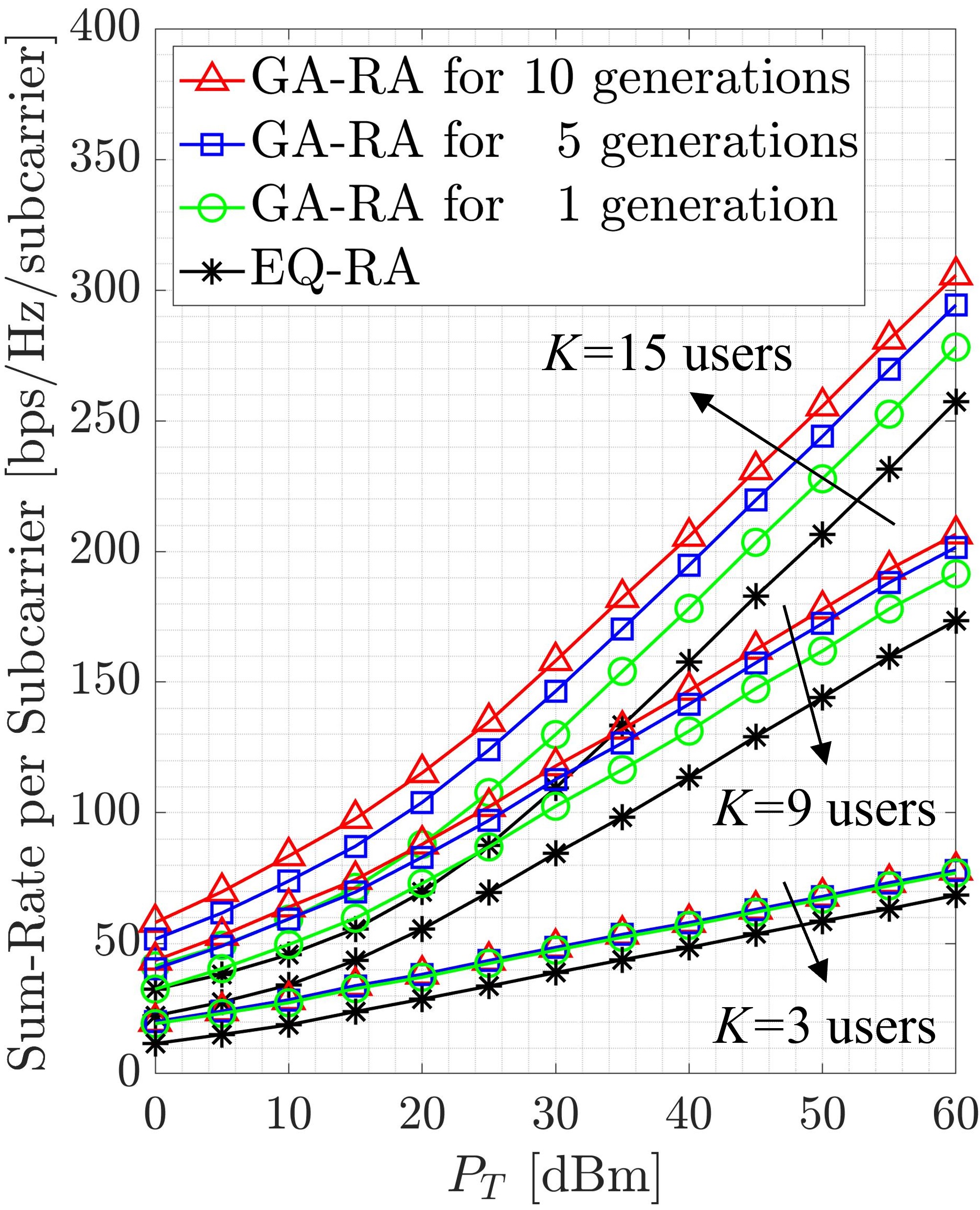}
		\caption{$C = 8$ subcarriers}
		\label{fig: SR per sb vs pt 8}
	\end{subfigure}
	\hfill
	\begin{subfigure}{.24\textwidth}
		\centering
		\includegraphics[width = \textwidth]{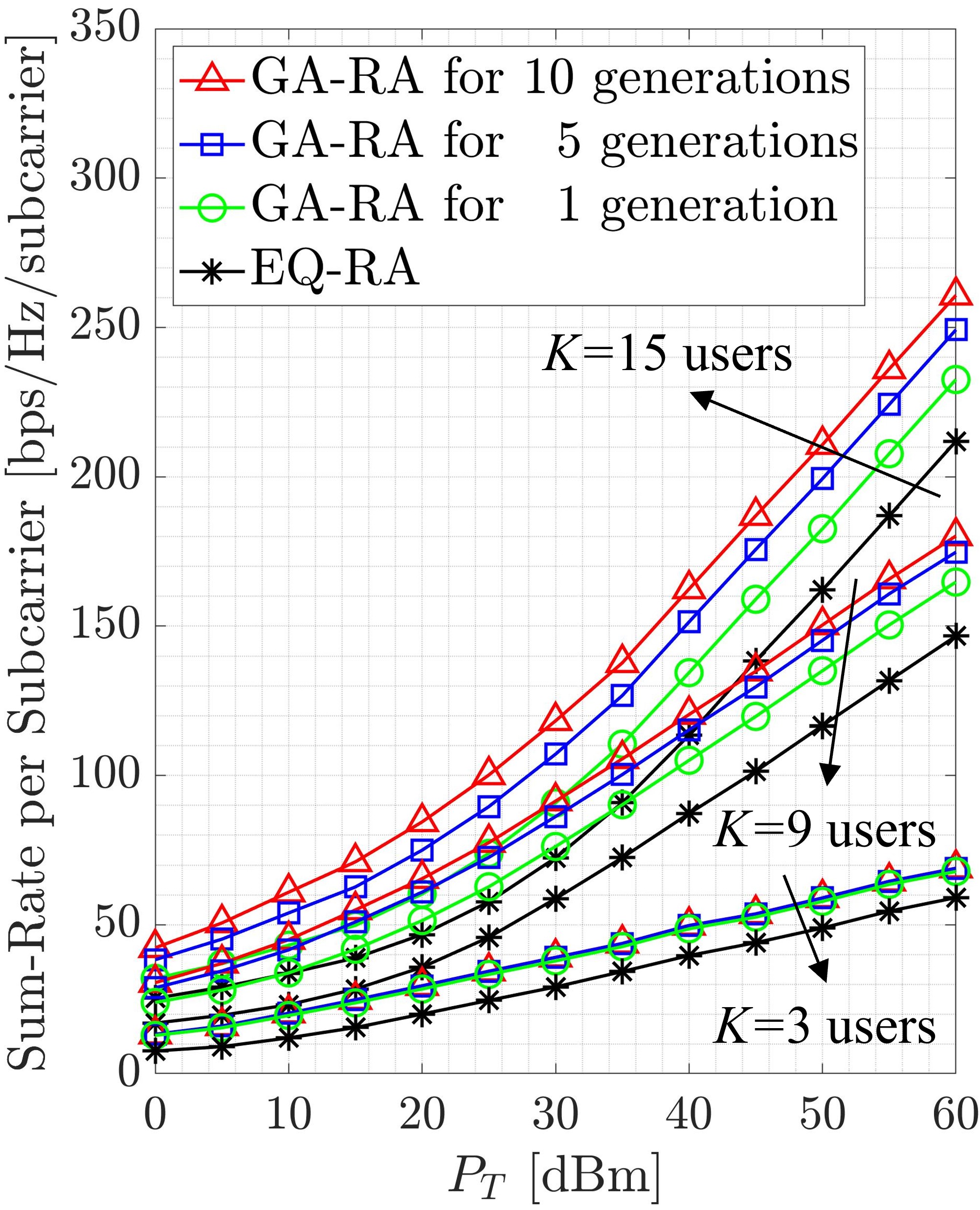}
		\caption{$C = 64$ subcarriers}
		\label{fig: SR per sb vs pt 64}
	\end{subfigure}
	\vspace{-0.1ex}
	\caption{Sum-rate per subcarrier  versus the transmitted power. \label{fig: sum-rate}}
\end{figure}

The average sum-rate per subcarrier is plotted versus the total transmit power for $C = 8$ and $C = 64$ subcarriers in Fig. \ref{fig: sum-rate}, where $K = 3, 9, 15$ downlink users are served.
It is observed that the proposed GA-RA technique significantly enhances the system capacity. For example, when the transmit power is $P_T=40$ dBm, applying the proposed GA-RA technique with $Q=10$ generations enhances the sum-rate performance for $C=8$ subcarriers by $19.66\%$, $29.4\%$, $30.5\%$ and for $C=64$ subcarriers by $25.41\%$, $38\%$, $43.06\%$ compared with EQ-RA technique for $K = 3, 9, 15$ users, respectively. In addition, the performance gap between GA-RA and EQ-RA increases for the larger number of users.
On the other hand, when there are less number of users (e.g., $K=3$ users), the sum-rate performance is saturated in earlier generations.

\begin{figure}[!t]
	\begin{subfigure}{.48\columnwidth}
		\centering
		\includegraphics[width = \textwidth]{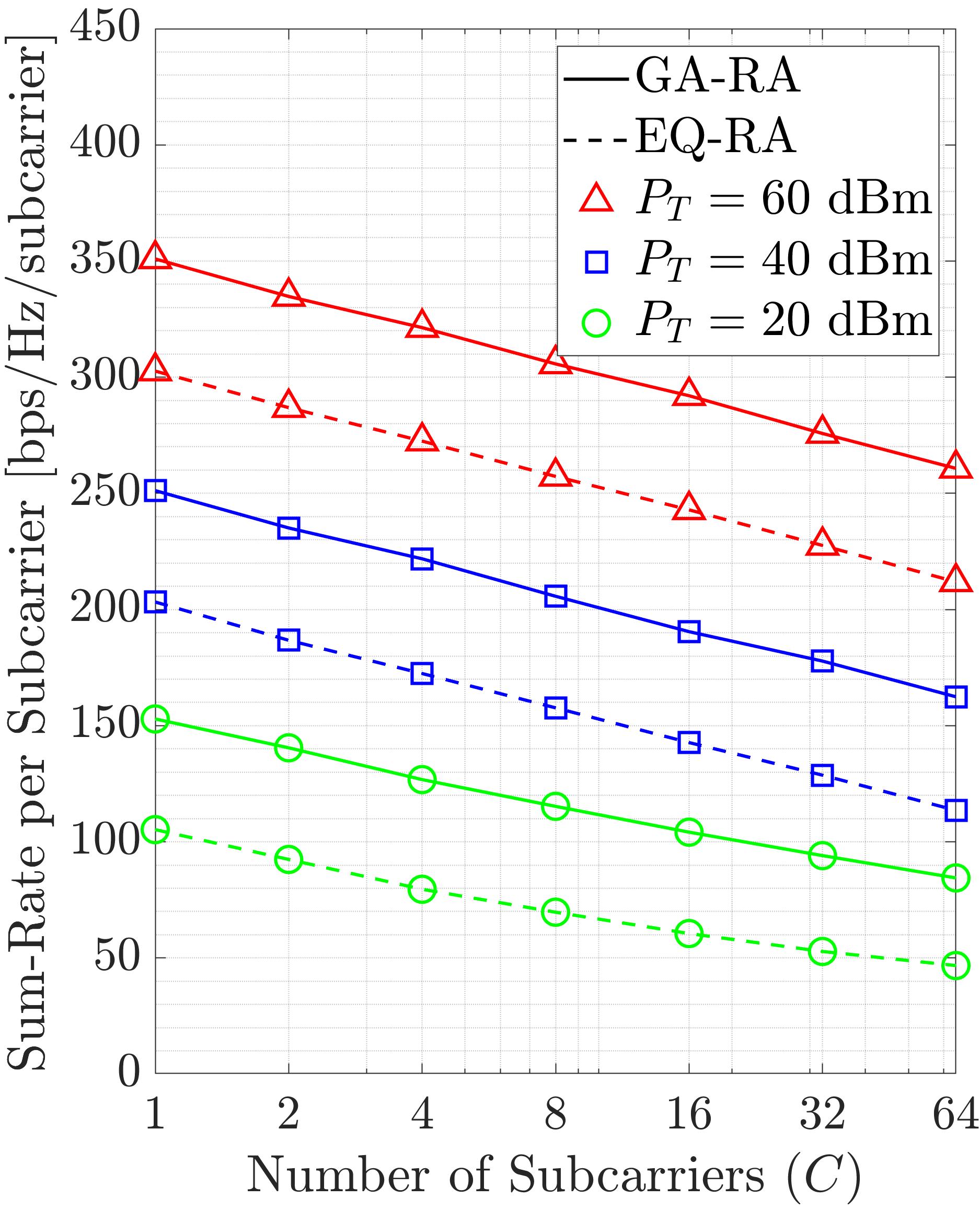}
		\caption{}
		\label{fig: SumRate_vs_subcarrierNumber}
	\end{subfigure}
	\hfill
	\begin{subfigure}{.49\columnwidth}
		\includegraphics[width = \textwidth]{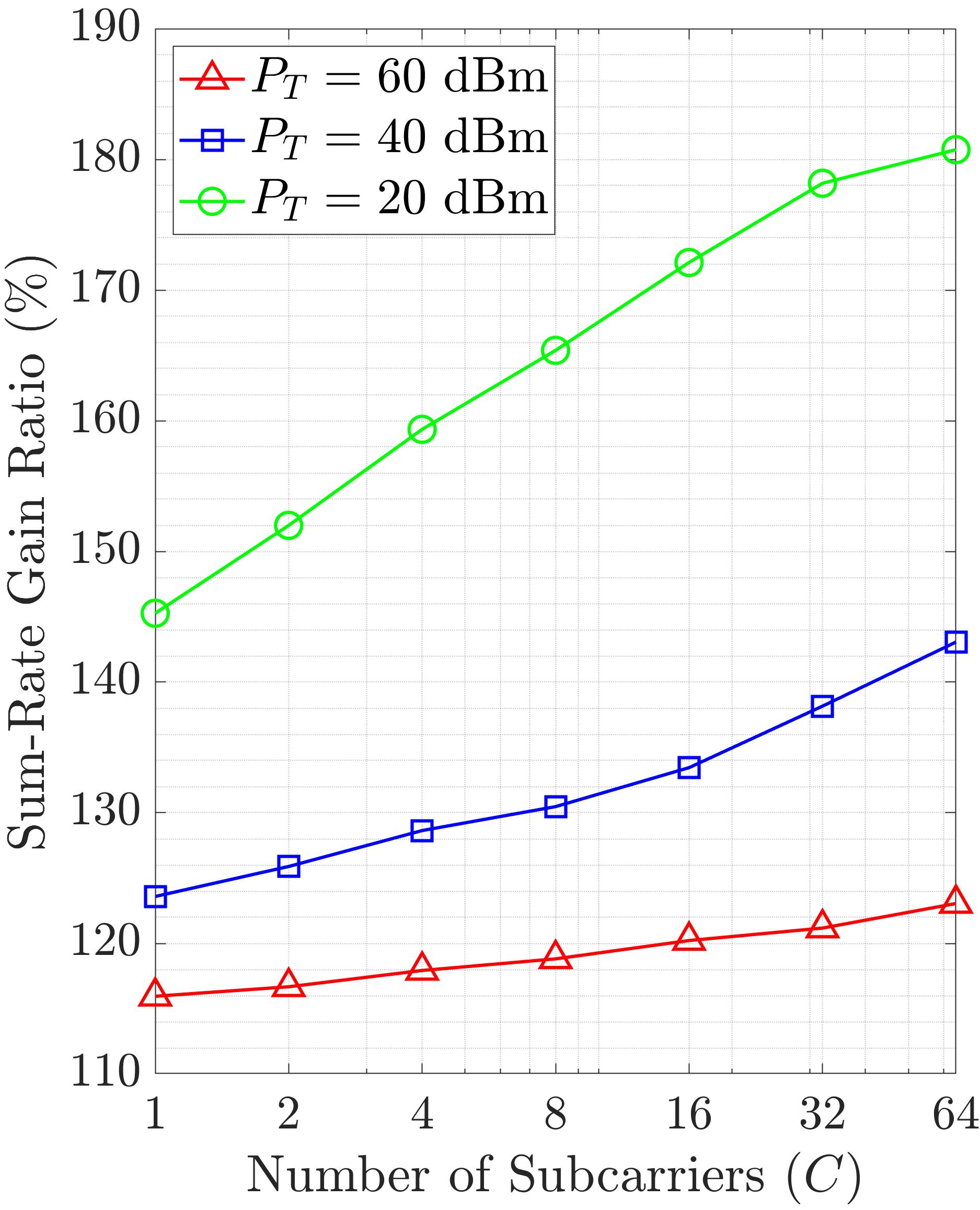}
		\caption{}
		\label{fig: SumRateGR_vs_subcarrierNumber}
	\end{subfigure}
	\caption{ (a) Sum-rate per subcarrier and (b) sum-rate gain ratio versus number of subcarriers ($K=15$ users).} \label{fig: S_and_SGR}
	\vspace{-2ex}
\end{figure}

Fig. \ref{fig: S_and_SGR}(a) compares the average sum-rate per subcarrier performance of GA-RA and EQ-RA techniques versus the number of subcarriers, where the transmit power is chosen as $P_T = 20 , 40, 60 \textrm{ dBm}$. Here, we observe that the sum-rate improvement provided by the GA-RA technique is almost constant for different subcarrier scenarios. Specifically, the performance of GA-RA is better than EQ-RA by approximately $48.5$ bps/Hz/subcarrier at $P_T = 40, 60$ dBm. However, at $P_T=20$ dBm, the performance improvement is $47.6$ bps/Hz/subcarrier for $C=1$ subcarrier and it decays $38.2$ bps/Hz/subcarrier for $C=64$ subcarriers. 
Moreover, it is shown that by increasing the number of subcarriers, the sum-rate per subcarrier performance degrades since less power is allocated to each subcarrier as discussed in expressed in \eqref{eqn: Optimization}. 
	
By defining the sum-rate gain ratio as $\frac{R_{sum,GA}^{(Q)}}{R_{sum,EQ}}$, Fig.  \ref{fig: S_and_SGR}(b) depicts that the proposed GA-RA technique enhances further the sum-rate gain ratio by increasing the number of subcarriers. The importance of this fact appears is more apparent in the case of low transmitted power.

	
\section{Conclusions}\label{SEC_CONC}

In this work, a novel genetic algorithm based resource allocation (GA-RA) technique has been developed for {energy-efficient throughput maximization} in the MU-mMIMO-OFDM systems. Furthermore, the angular-based hybrid precoding (AB-HP) scheme has been developed for the OFDM-based downlink transmission to reduce the number of RF chains and the CSI overhead size. The AB-HP scheme first develops a single RF beamformer block for all subcarriers via AoD information, then builds a distinct BB precoder for each subcarrier via the reduced-size effective CSI.
Next, the GA-RA technique allocates power and subcarrier resources among the users.
The promising numerical results demonstrate that the GA-RA technique greatly enhances the sum-rate (i.e., {throughput}) performance of MU-mMIMO-OFDM systems compared to the conventional equal RA (EQ-RA). Also, the performance gain ratio with GA-RA increases for the larger number of subcarriers, especially in case of low transmission power. 

\ifCLASSOPTIONcaptionsoff
\newpage
\fi
\bibliographystyle{IEEEtran}
\bibliography{bibAsil_2112}

\end{document}